\documentclass[amsmath,amssymb,showpacs,twocolumn]{revtex4}
\usepackage{graphicx}
\usepackage{dcolumn}
\usepackage{bm}
\makeatletter \@addtoreset{equation}{section} \makeatother

\let\^=\hat

\begin{document}
\title{Fully Localized Two-dimensional Embedded Solitons}
\author{Jianke Yang}
\affiliation{Department of Mathematics and Statistics, University of
Vermont, Burlington, VT 05401, USA}

\begin{abstract}
We report the first prediction of fully localized two-dimensional
embedded solitons. These solitons are obtained in a
quasi-one-dimensional waveguide array which is periodic along one
spatial direction and localized along the orthogonal direction.
Under appropriate nonlinearity, these solitons are found to exist
inside the Bloch bands (continuous spectrum) of the waveguide, and
thus are embedded solitons. These embedded solitons are fully
localized along both spatial directions. In addition, they are fully
stable under perturbations. These results show that multidimensional
embedded solitons may find applications just like non-embedded
(regular) solitons.
\end{abstract}

\maketitle

Embedded solitons are nonlinear solitary waves whose frequencies (or
propagation constants) reside inside the continuous spectrum of the
underlying wave system \cite{Yang_PRL99}. The existence of embedded
solitons is quite counter-intuitive, since inside the continuous
spectrum, only nonlocal waves with nonvanishing oscillating tails
are commonly expected \cite{Boyd_book}. However, under certain
conditions, these oscillating tails are absent, hence truly
localized embedded solitons appear inside the continuous spectrum
\cite{Yang_PRL99,Champneys_ES}. Since embedded solitons exist inside
the continuous spectrum and are thus resonant with linear radiation
modes, they exhibit some interesting dynamical properties. For
instance, isolated embedded solitons are often found to be
semi-stable, i.e., they would persist under energy-enhancing
perturbations but perish under energy-reducing perturbations
\cite{Yang_PRL99,Tan_ES}. Non-isolated embedded solitons, on the
other hand, can be semi-stable or fully stable, depending on the
underlying wave system \cite{Yang_Akylas_ES,Yang_PRL03}. Embedded
solitons are linked to other physical objects as well. For instance,
moving discrete solitary waves in lattices (if they exist) are also
embedded solitons
\cite{Peli_discrete,Malomed_discrete,Champneys_lattice}. So far, all
embedded solitons reported in the literature are one-dimensional
(1D) to the author's knowledge. The soliton trains reported in
\cite{ES_train} exist inside the continuous spectrum and are
two-dimensional (2D), but these soliton trains are localized only
along one spatial direction and nonlocal along the orthogonal
direction. It has remained a challenge to find 2D (and
higher-dimensional) embedded solitons which are localized in all
spatial directions. The reason is that in multi-dimensions, more
stringent conditions need to be satisfied in order for fully
localized embedded solitons to exist, thus such solitons are more
difficult to find.

From a broader perspective, embedded solitons are intimately related
to linear bound states (i.e., localized eigenmodes) inside the
continuous spectrum of a wave system. These linear bound states in
the continuum were first predicted by von Neumann and Wigner in 1929
\cite{Newmann}, who showed that the 3D linear Schr\"odinger equation
with certain localized potentials could possess bound states above
the potential well (see also \cite{Stillinger_1975}). These
predicted bound states were later observed experimentally for
electrons in semiconductor heterostructures \cite{Capasso_1992}. In
optics, linear bound states inside the continuum have also been
predicted in various settings, such as a semi-infinite 1D lattice
\cite{Longhi_2007}, two parallel dielectric gratings
\cite{Marinica_2008}, two arrays of thin parallel dielectric
cylinders \cite{Marinica_2008}, and open 2D quantum dots or optical
waveguides \cite{Sadreev_2006,Moiseyev_2009}. Recently, linear 2D
bound states in the continuum were demonstrated both theoretically
and experimentally for light beams in a quasi-1D waveguide array
with two additional waveguides above and below it \cite{Segev}. The
key idea in the construction of linear continuum bound states in
\cite{Sadreev_2006,Moiseyev_2009,Segev} is to seek bound states of
certain parity which are embedded inside the continuum bands of
opposite parity. This idea inspires us to construct fully localized
2D embedded solitons in this paper. The above theoretical and
experimental investigations on the counter-intuitive nonlinear
embedded solitons and linear continuum bound states deepened our
fundamental understanding of linear and nonlinear wave phenomena,
and they could lead to unexpected applications in diverse physical
fields.

In this paper, we construct fully localized 2D embedded solitons for
the first time. These solitons are obtained in a quasi-1D waveguide
array which is periodic along the horizontal direction and localized
along the vertical direction. Under self-defocusing nonlinearity, we
find 2D solitons which are symmetric along the vertical direction,
and they are embedded in the continuum bands of odd symmetry in the
vertical direction. These 2D embedded solitons exist as continuous
families, with their propagation constants (or equivalently their
powers) as a free parameter. We further show that these embedded
solitons are fully stable against perturbations even though they
exist inside the continuum bands. In addition, we show how 2D
embedded solitons of odd symmetry along the vertical direction can
be derived under self-focusing nonlinearity. This construction
method for 2D embedded solitons is general, thus these embedded
solitons are not rare objects, but can appear easily in diverse
physical situations.

The theoretical model we use is the following 2D NLS equation with a
potential,
\begin{equation} \label{NLS}
iU_z+U_{xx}+U_{yy}+n(x,y)U+\sigma |U|^2U=0.
\end{equation}
In spatial optics, this equation models paraxial light transmission
in a waveguide under cubic nonlinearity \cite{Maxim_book}. In this
context, $U$ is the complex envelope function of the light's
electric field, $z$ is the transmission distance, $(x,y)$ are the
transverse coordinates, $n(x, y)$ is the refractive index variation
of the waveguide, and $\sigma=\pm 1$ represent self-focusing and
self-defocusing nonlinearity respectively (self-focusing
nonlinearity is common in most optical materials, and
self-defocusing nonlinearity can be realized in certain special
materials such as photorefractive crystals \cite{Segev_nature}). In
Bose-Einstein condensates, Eq. (\ref{NLS}) models the collective
behavior of condensate atoms in a magnetic or optical trap under
nonlinear atom-atom interaction (it is called the Gross-Pitaevskii
equation in the literature) \cite{GP}.
Static solitary waves in Eq. (\ref{NLS}) are sought in the form
\begin{equation}
U(x,y,z)=u(x,y)e^{-i\mu z},
\end{equation}
where $\mu$ is the propagation constant, and $u(x,y)$ is a
real-valued localized function which satisfies the equation
\begin{equation} \label{ODE}
u_{xx}+u_{yy}+n(x,y)u+\sigma u^3=-\mu \hspace{0.02cm} u.
\end{equation}
To construct a concrete example of 2D embedded solitons, we take a
quasi-1D waveguide array
\begin{equation} \label{nxy}
n(x,y)=6\cos^2\hspace{-0.06cm}x \hspace{0.06cm} e^{-y^2/4},
\end{equation}
which is periodic along the $x$-direction and localized along the
$y$-direction. This waveguide is shown in Fig. 1(left panel). We
also take $\sigma=-1$ (for self-defocusing nonlinearity). In order
to find embedded solitons, we first need to determine the linear
continuous spectrum of Eq. (\ref{ODE}). For this purpose, we drop
the nonlinear term in (\ref{ODE}). Since $n(x,y)$ is periodic in $x$
with $\pi$ period, according to the Bloch theorem, linear eigenmodes
of (\ref{ODE}) are of the form
\begin{equation} \label{uBloch}
u(x, y)=e^{ikx}q(x,y),
\end{equation}
where $k$ is the wavenumber in the first Brillouin zone $-1\le k\le
1$, and $q(x,y)$ is an $x$-periodic function with period $\pi$.
The continuous spectrum of Eq. (\ref{ODE}) consists of the positive
axis $\mu\in [0, +\infty)$, where $u(x,y)$ is nonlocalized in the
$y$ direction, and Bloch bands with $\mu <0$, where $u(x,y)$ is
localized in the $y$ direction. To determine the Bloch bands with
$\mu <0$, we expand $q(x,y)$ into Fourier series in both $x$ and
$y$, then insert (\ref{uBloch}) into the linear part of Eq.
(\ref{ODE}) and turn it into a matrix eigenvalue problem, with $\mu$
being the eigenvalue and the Fourier coefficients of $q(x,y)$ being
the eigenvector \cite{Maxim_book}. This matrix eigenvalue problem is
then solved by conventional algorithms. The resulting diffraction
relation $\mu=\mu(k)$ for these Bloch bands is shown in Fig. 1(right
panel). We see that three Bloch bands are obtained. At the edges of
the lowest two bands, the corresponding Bloch modes $u(x,y)$ are
displayed in Fig. 2. It is important to notice that the Bloch modes
in the lowest band $\mu\in [-2.9711, -2.7226]$ are symmetric in $y$,
while the Bloch modes in the second band $\mu \in [-1.3030,
-0.9260]$ are anti-symmetric in $y$. The existence of different
Bloch bands with opposite $y$-parity is important for our
construction of 2D embedded solitons.

\begin{figure}
\includegraphics[width=0.45\textwidth]{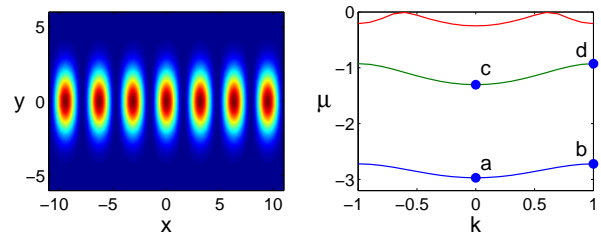}
\caption{Left: the quasi-1D waveguide array $n(x,y)$ in our model.
Right: the Bloch bands in this waveguide. The edges of the lowest
two Bloch bands are marked by letters `a-d'. \label{fig1} }
\end{figure}

\begin{figure}
\includegraphics[width=0.45\textwidth]{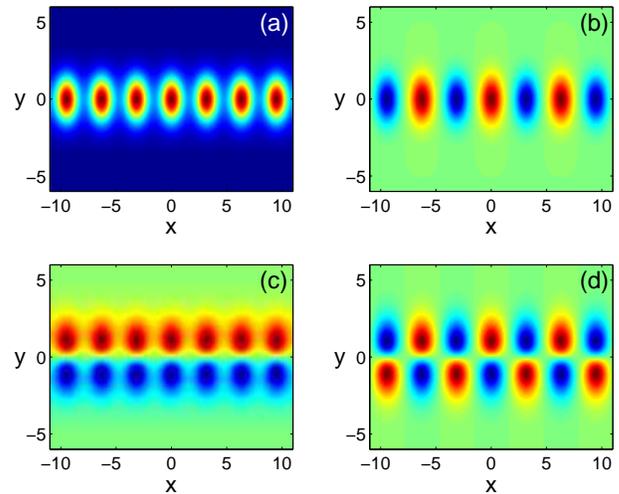}
\caption{Bloch modes at edges `a-d' of the lowest two Bloch bands in
Fig. 1 respectively.    \label{fig2} }
\end{figure}

When nonlinearity is present, locally-confined solitons will
bifurcate out from infinitesimal (linear) Bloch modes of band edges
\cite{Peli_PRE,Maxim_book}. Under self-defocusing nonlinearity,
these solitons will bifurcate from the upper band edges upward into
band gaps. Here we consider the solitons bifurcating from the upper
edge of the lowest Bloch band (i.e., point `b' in Fig. 1). Near edge
`b', the soliton is a low-amplitude broad packet which decays slowly
along the $x$-direction (see Fig. 3(A)). This soliton is a regular
gap soliton since it exists in a band gap. As $\mu$ moves further
away from the edge `b', the soliton becomes more narrow, and its
amplitude as well as power becomes higher (see Fig. 3). Here the
power $P$ is defined as the integral of $u^2$ over the $(x,y)$
plane. The most interesting phenomenon about this family of solitons
is that, when $\mu$ enters into the second Bloch band $[-1.3030,
-0.9260]$, the soliton still persists, and it remains fully
localized in both $x$ and $y$ directions. To demonstrate, this
soliton at $\mu=-1.1$ in the middle of the second Bloch band is
displayed in Fig. 3(B). Since this soliton exists inside the
continuous spectrum (Bloch bands), it is a fully-localized 2D
embedded soliton! Likewise, its nearby solitons with $\mu$ still
inside the second Bloch band are all 2D embedded solitons as well.
In other words, this is a continuous family of 2D embedded solitons
with its propagation constant $\mu$ or power $P$ as a free
parameter. This is the first report of 2D embedded solitons to our
best knowledge.

\begin{figure}
\includegraphics[width=0.45\textwidth]{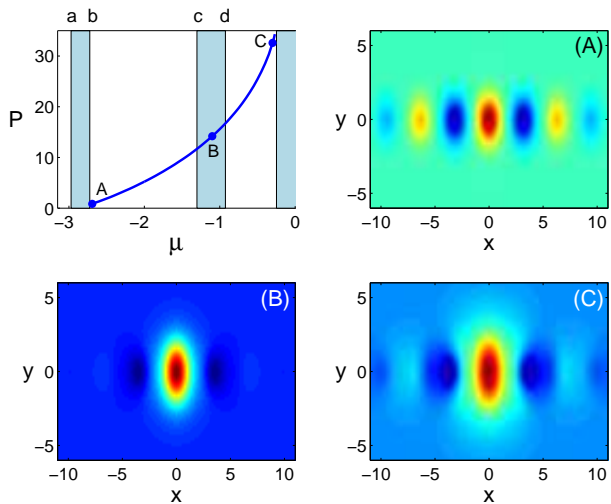}
\caption{A soliton family under self-defocusing nonlinearity
($\sigma=-1$). Upper left panel: the power curve;  the shaded
stripes indicate Bloch bands, and the band edges marked by `a-d' on
top of the panel correspond to the edges of the same marker in Fig.
1. Panels (A-C) are soliton profiles $u(x,y)$ at locations marked by
the same letters on the power curve. The soliton at location `B' is
an embedded soliton which is fully localized in both $x$ and $y$
dimensions. \label{fig3} }
\end{figure}

Why do these 2D embedded solitons exist? Notice that these solitons
bifurcate out from edge `b' of the first Bloch band, thus they are
symmetric in $y$ (see Fig. 3). Notice also that the second Bloch
band consists of Bloch modes which are all anti-symmetric in $y$
(see Fig. 2). Thus when this $y$-symmetric soliton branch enters the
second Bloch band of $y$-antisymmetric Bloch modes, even though
$\mu$ lies in the Bloch band, the soliton does not excite those
Bloch modes of opposite $y$-parity, thus it remains fully localized.
However, if this soliton moves into the third band $\mu\in [-0.2488,
0]$ (see Fig. 3), since the Bloch modes in this third band are also
symmetric in $y$, this soliton \emph{will} excite these
$y$-symmetric Bloch modes and become delocalized (evidence of this
can be seen in Fig. 3(C), where the soliton becomes broad again near
the third Bloch band). Thus one can not find $y$-symmetric 2D
embedded solitons in the third band.

Stability of these 2D embedded solitons in Fig. 3 is an important
issue. In previous studies of 1D embedded solitons, since the
embedded soliton is in resonance with the continuous spectrum, it
could excite the continuum radiation and perish under certain
perturbations \cite{Yang_PRL99}. Those 1D embedded solitons could
also be linearly unstable, leading to their destruction under any
generic perturbation \cite{Champneys_ES}. For the 2D embedded
solitons in Fig. 3, we have found that they are all linearly stable,
i.e., their linear-stability spectra do not contain any eigenvalues
with positive real parts. This linear stability for the embedded
soliton in Fig. 3(B) is demonstrated in Fig. 4(i). Regarding the
question of nonlinear stability, we note that these 2D embedded
solitons lie inside the second Bloch band whose Bloch modes are
antisymmetric in $y$. Thus if the perturbation is $y$-symmetric as
the embedded soliton itself, then since the waveguide $n(x,y)$ is
also $y$-symmetric, the solution of Eq. (\ref{NLS}) will remain
$y$-symmetric for all distances $z$. Hence the perturbed soliton
would not excite $y$-antisymmetric second-band modes, i.e., the
soliton would be stable under $y$-symmetric perturbations. A less
trivial question is what would happen if the perturbation is
asymmetric in $y$. In this case, the perturbed soliton would excite
$y$-antisymmetric second-band modes since it is resonant with those
modes. Then could these $y$-antisymmetric radiation break up the
embedded soliton? Intuitively, we can expect that when the
$y$-antisymmetric component of the perturbation is weak, then these
weak antisymmetric components would disperse away through resonance
with the second-band modes, and the other dominant $y$-symmetric
component of the solution would adjust its shape into a nearby
($y$-symmetric) embedded soliton. If so, then these 2D embedded
solitons would be nonlinearly fully stable. However, this
expectation is under the assumption that energy in the $y$-symmetric
component would not transfer to the $y$-antisymmetric component
during evolution. Since Eq. (\ref{NLS}) is nonlinear, this
assumption may not hold, because the symmetric and antisymmetric
components could couple each other and transfer energy between them.
Thus in principle, it is possible for the perturbed soliton to lose
a significant amount of radiation to the resonant second-band modes
and break up. To clarify this question, we have performed numerical
simulations of these embedded solitons under various asymmetric
perturbations, and found that they are always nonlinearly fully
stable. Two typical simulation results are shown in Fig. 4(ii-iv).
In these simulations, the embedded soliton $u(x,y)$ is the one in
Fig. 3(B), and the perturbed initial state is
\begin{equation} \label{ic}
U(x,y,0)=u(x,y)+\epsilon \hspace{0.04cm} (1+\sin y) \hspace{0.04cm}
e^{-(x^2+y^2)/4},
\end{equation}
where $\epsilon$ is the strength of perturbations. Notice that this
perturbation contains both symmetric and anti-symmetric components
in $y$. For $\epsilon=0.2$, this perturbed initial state is shown in
Fig. 4(ii). At propagation distance $z=50$, the solution is shown in
Fig. 4(iii). It is seen that this embedded soliton is stable under
this perturbation. This stability can be seen more clearly in Fig.
4(iv), where the peak amplitude $|U|_{\mbox{max}}$ of the solution
versus the propagation distance $z$ is displayed. We can see that
the peak amplitude approaches a constant value close to the
amplitude of the unperturbed soliton at large distances. If we take
a different perturbation with $\epsilon=-0.2$, the result is
similar, i.e., the peak amplitude of the solution also approaches a
constant value close to the amplitude of the unperturbed soliton at
large distances (see Fig. 4(iv)). Thus this embedded soliton is
nonlinearly fully stable. This result resembles the full stability
of 1D embedded solitons in a generalized third-order NLS equation
\cite{Yang_PRL03}. It contrasts some other 1D embedded solitons
which are semi-stable and perish under certain types of
perturbations \cite{Yang_PRL99,Yang_Akylas_ES}.

\begin{figure}
\includegraphics[width=0.45\textwidth]{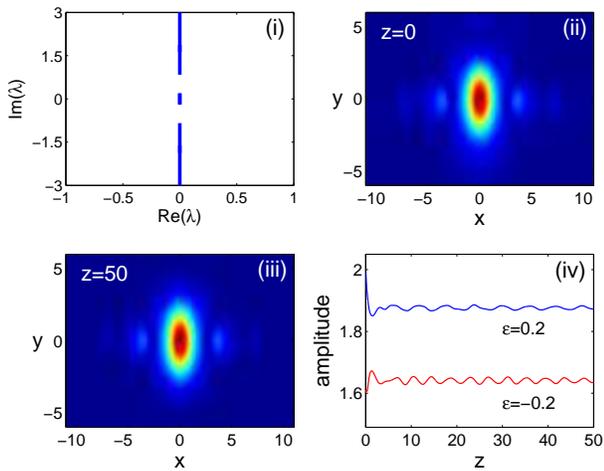}
\caption{Demonstration of stability for the embedded soliton in Fig.
3(B). (i) the linear-stability spectrum, showing that this soliton
is linearly stable; (ii) an initially perturbed embedded soliton
(\ref{ic}) with $\epsilon=0.2$; (iii) evolution of the perturbed
soliton in (ii) at $z=50$; plotted in (ii, iii) are $|U|$ fields;
(iv) peak-amplitude evolutions of the perturbed embedded soliton
(\ref{ic}) for $\epsilon=0.2$ and $-0.2$. \label{fig4} }
\end{figure}

\begin{figure}[t]
\includegraphics[width=0.45\textwidth]{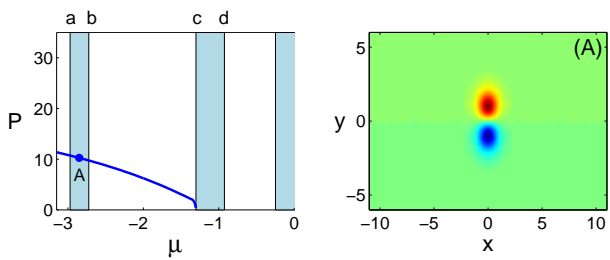}
\caption{2D embedded solitons under self-focusing nonlinearity
($\sigma=1$). Left panel: the power curve of this soliton family
(same notations as Fig. 3); right panel: an embedded soliton in the
first Bloch band (at location `A' of the power curve). \label{fig5}
}
\end{figure}

It is important to recognize that the method in this paper for the
construction of 2D embedded solitons is quite general, and it can be
used to construct many other fully localized multidimensional
embedded solitons in various physical systems. For instance, in the
same quasi-1D waveguide array described by Eqs. (\ref{NLS}) and
(\ref{nxy}), if self-focusing nonlinearity is taken (i.e.,
$\sigma=1$), then a soliton family which is anti-symmetric in $y$
will bifurcate downward from edge `c' of the second Bloch band (see
Figs. 1 and 2). This solution family passes through the first Bloch
band whose Bloch modes are symmetric in $y$. Inside the first Bloch
band, these solitons are also fully-localized 2D embedded solitons.
To demonstrate, the power curve of this soliton family is displayed
in Fig. 5 (left panel). At point `A' inside the first Bloch band,
the embedded soliton is shown in Fig. 5(A), which is fully localized
in both dimensions. Using similar methods, we can construct fully
localized 3D embedded solitons as well. Previously, embedded
solitons were generally regarded as rare objects which appear by
``accident". Now we see that embedded solitons can arise frequently
in diverse physical situations, thus they are an important physical
object in nonlinear wave systems.

Now we address why the above results are of interest to physics and
mathematics. Intuitively, solitary waves are only expected outside
the continuous spectrum. In the past ten years, the counterintuitive
concept of solitons embedded inside the continuous spectrum was
proposed and demonstrated in one dimension
\cite{Yang_PRL99,Champneys_ES}. This concept significantly deepened
our general understanding of nonlinear wave phenomena. In addition,
it fostered the construction of other physical 1D objects such as
moving discrete solitons in lattices since those objects are also
embedded solitons under disguise
\cite{Peli_discrete,Malomed_discrete,Champneys_lattice}. In this
paper, for the first time to our knowledge, we demonstrated the
existence of embedded solitons in two dimensions, and pointed out a
way to construct embedded solitons in even higher dimensions (such
as three dimensions). This significantly broadened the scope of
embedded solitons. It could also stimulate the construction of
related objects such as multidimensional moving discrete solitons.
From the viewpoint of physical applications, regular (non-embedded)
solitons have found applications in numerous situations. Now with
the demonstration of families of stable multidimensional embedded
solitons in this paper, these embedded solitons may find
applications analogous to regular solitons in situations where
regular solitons do not exist.

In summary, we have predicted fully localized 2D embedded solitons
for the first time. These embedded solitons were obtained in a
quasi-1D waveguide array, and they exist inside the Bloch bands
whose Bloch modes have opposite parity from the solitons themselves.
These embedded solitons form solution families with continuous
ranges of power values. In addition, they are fully stable under
perturbations. The method of construction in this paper is general,
and it can be used to obtain multi-dimensional embedded solitons in
diverse physical systems.

This work was supported in part by AFOSR grant USAF 9550-09-1-0228
and NSF grant DMS-0908167.


\end{document}